\documentclass{scrartcl} 

\usepackage[utf8]{inputenc}
\usepackage{pgf, tikz}
\usetikzlibrary{arrows, automata}
\usetikzlibrary{positioning}
\usepackage{amsmath}
\usepackage{amssymb}
\usepackage{mathabx}
\usepackage{mathtools}

\usepackage[linesnumbered,ruled,vlined]{algorithm2e}
\usepackage{natbib}
\usepackage{dsfont}

\newcommand{\R}{\mathbb{R}} % set R
\newcommand{\setX}{\mathcal{X}} % set X (curly)
\newcommand{\bigO}{\mathcal{O}} % set X (curly)
\newcommand{\E}{\mathbb{E}} % set E
\newcommand{\Proba}{\mathbb{P}} % set P
\newcommand{\eqdef}{:=}
\newcommand{\Var}{\mathrm{Var}}
\newcommand{\Cov}{\mathrm{Cov}}
\newcommand{\sgn}{\mathrm{sgn}}
\newcommand{\ind}{\mathds{1}}
\newcommand{\footremember}[2]{%
   \footnote{#2}
    \newcounter{#1}
    \setcounter{#1}{\value{footnote}}%
}
\newcommand{\footrecall}[1]{%
    \footnotemark[\value{#1}]%
} 

\SetKwInput{KwInput}{Input}       
\SetKwInput{KwOutput}{Output}

\title{Fast compression of MCMC output}
\author{Nicolas Chopin\footnote{ENSAE, IPP; email: nicolas.chopin@ensae.fr} \footremember{equal}{These authors contributed equally to this work.}\hspace{0.3em} and Gabriel Ducrocq\footnote{ENSAE, IPP; email: gabriel.ducrocq@ensae.fr} \footrecall{equal}}

\begin{document}

\maketitle
%\firstnote{Current address: Affiliation 3} 
%\secondnote{These authors contributed equally to this work.}

\abstract{    
We propose cube thinning, a novel method for compressing the output of a MCMC (Markov chain Monte Carlo) algorithm when control variates are available. It amounts to resampling  the initial MCMC sample (according to weights derived from  control variates), while imposing equality constraints on averages of these control variates, using the cube method of \cite{Deville2004}. 
    Its main advantage is that its CPU cost is linear in $N$, the original sample size, and is constant in $M$, the required size for the compressed sample.  
    This compares favourably to  Stein thinning \citep{Riabiz2020}, which has complexity $\bigO(NM^2)$, and which requires the availability of the gradient of the target log-density (which automatically implies the availability of control variates). 
    Our numerical experiments suggest that cube thinning is also competitive in terms of statistical error. 
    }
\section{Introduction}

MCMC (Markov chain Monte Carlo) remains to this day the most popular approach to sampling from a target distribution $p$, in particular in Bayesian computation \citep{Robert2004}.

Standard practice is to run a single chain, $X_1,\ldots,X_N$ according to a Markov kernel that leaves invariant $p$. It is also common to discard part of the simulated chain, either to reduce its memory footprint, or to reduce the CPU cost of later post-processing operations, or more generally for the user's convenience. 
Historically, the two common recipes for compressing MCMC output are: 
\begin{itemize}
    \item burn-in, which amounts to discarding the $b$ first states; and
    
    \item thinning, which amounts to retaining only one out of $t$ (post burn-in) states. 
\end{itemize}

The impact of either recipes on the statistical properties of the sub-sampled estimates are markedly different. Burn-in reduces the bias introduced  by the discrepancy between $p$ and the  distribution of the initial state $X_1$ (since $X_b\approx p$ for $b$ large enough). On the other hand, thinning always increases the (asymptotic) variance of MCMC estimates \citep{Geyer1992}.

Practitioners often choose $b$ (the burn-in period) and $t$ (the thinning frequency) separately, in a somewhat ad-hoc fashion (i.e. through visual inspection of the initial chain), or using convergence diagnosis such as e.g. those reviewed in \cite{MR1395755}.
 
Two recent papers \citep{Mak2018, Riabiz2020}, cast a new light on the problem of compressing a MCMC chain by considering more generally the problem, for a given $M$, of  selecting the subsample of size $M$ that best represents (according to a certain criterion) the target distribution $p$. 
We focus for now on \cite{Riabiz2020}, for reasons we explain below.

Stein thinning, the method developed in \cite{Riabiz2020}, 
chooses the sub-sample $\mathcal{S}$ of size $M$ which minimises the following criterion: 
\begin{equation}
    \label{eq:ksd}
D(\mathcal{S}) \eqdef \frac 1 {M^2} \sum_{m, n\in \mathcal{S}} k_p(X_m, X_n),
\quad \mathcal{S}\subset \{1,\ldots, N\},\quad |\mathcal{S}|=M
\end{equation}
where $k_p$ is a $p-$dependent kernel function derived from another kernel function $k:\setX\times\setX \rightarrow \R$, as follows: 
\begin{multline*}
k_p(x, y) = \nabla_x \cdot \nabla_y k(x, y) 
+ \langle \nabla_x k(x, y), s_p(y)\rangle \\
+ \langle \nabla_y k(x, y) , s_p(x)\rangle 
+ k(x, y) \langle s_p(x), s_p(y)\rangle 
\end{multline*}
with $\langle \cdot, \cdot\rangle$ being the Euclidean inner product, 
$s_p(x)\eqdef \nabla\log p(x)$ is the so-called score function 
(gradient of the log target density), and $\nabla$ the gradient operator. 

The rationale behind criterion \eqref{eq:ksd} is that it may be interpreted as the KSD (kernel Stein divergence) between the true distribution $p$ and the empirical distribution of sub-sample $S$.  
We refer to  \cite{Riabiz2020} for more details on the theoretical background of the KSD, and its connection to Stein's method. 

Stein thinning is appealing, as it seems to offer a principled, quasi-automatic way to compress MCMC output. However, closer inspection reveals the following three limitations. 

First, it requires computing the gradient of the log-target density, 
$s_p(x)=\nabla\log p(x)$. 
This restricts the  method to problems where this gradient exists and is tractable (and, in particular, to $\setX=\R^d$). 

Second, its CPU cost is $\bigO(NM^2)$. This makes it nearly impossible to use Stein thinning for $M\gg 100$. This cost stems from the
greedy algorithm proposed in \cite{Riabiz2020}, see their Algorithm 1, which
adds at iteration $t$ the state $X_i$ which minimises $k_p(X_i, X_i) +
\sum_{j\in S_{t-1}} k_p(X_i, X_j)$, where $S_{t-1}$ is the sample obtained
from the $t-1$ previous iterations. 

Third, its performance seems to depend in a non-trivial way on the original kernel
function $k$; \cite{Riabiz2020} propose several strategies for choosing and scaling $k$, but none of them seems to perform uniformly well in their numerical experiments.

We propose a different approach in this paper, which we call cube thinning,
and which addresses these shortcomings to some extent.  Assuming
the availability of $J$ control variates (that is, of functions $h_j$ with known
expectation under $p$), we cast the problem of MCMC compression as that of
resampling the initial chain under constraints based on these control variates.
The main advantage of cube thinning is that its  complexity is $\bigO(NJ^3)$; in particular
it does not depend on $M$. That makes it possible to use it for much larger values of $M$. 
(We shall discuss the choice of $J$, but, by and large, $J$ should be of the same order as $d$, the dimension of the sampling space). 
The name stems from the cube method of \cite{Deville2004}, which plays a central part in our approach, 
as we explain in the body of the paper. 

The availability of control variates may seem like a strong requirement.
However, if we assume we are able to compute $s_p(x)=\nabla\log p(x)$, then  
(for a large class of functions $\phi:\R^{d}\rightarrow\R^{d}$, which we define later)
\[
    \E_p\left[ \phi(x) s_p(x) + \nabla_x \cdot \phi(x)   \right] = 0
\]
where $\nabla_x \cdot \phi$ denotes the divergence of $\phi$. 
In other words, the availability of the score function implies automatically the availability of  control variates. The converse is not true: there exists control variates \citep[e.g.][]{Dellaportas2011} that are not gradient-based. One of the examples we consider in our numerical examples feature such non gradient-based control variates; as a result, we are able to apply cube thinning, although Stein thinning is not applicable. 

The support point methods of \cite{Mak2018} does not require control variates. It is thus more generally applicable than either cube thinning or Stein thinning. On the other hand, when gradients (and thus control variates) are available, the numerical experiments of 
\cite{Riabiz2020} suggest that Stein thinning outperforms support points. 
From now on, we focus on situations where control variates are available.

The paper is organised as follows. Section~\ref{sec:steinCV} recalls the concept of
control variates, and explains how control variates may be used to reweight a
MCMC sample. 
Section~\ref{sec:cube} describes the cube method of \cite{Deville2004}. 
Section~\ref{sec:cubeThin} explains how to combine control variates and the cube method to 
perform cube thinning. 
Section~\ref{sec:experiments} assesses the statistical performance of cube thinning through
two numerical experiments. 

We use the following notations throughout: $p$ denotes both the target distribution and its probability density; $p(f)$ is a short-hand for the expectation of $f(X)$ under $p$. The gradient of a function $f$ is denoted by $\nabla_x f(x)$, or simply $\nabla f(x)$
when there is no ambiguity. 
The $i-$th component of a vector $v\in\R^d$ is denoted by $v[i]$, and its transpose by 
by $v^t$. The vectors of the canonical basis of $\R^d$ are denoted by $e_i$, i.e. $e_i[j]=1$ if $j=i$, 0 otherwise. Matrices are written in upper-case; the kernel (null space) of matrix $A$ is denoted by $\mathrm{ker} A$. 
The set of functions $f:\Omega\rightarrow \R^d$ that are continuously differentiable is denoted by $C^1(\omega, \R^d)$. 

\section{Control variates}\label{sec:controlVariates}

\subsection{Definition}

Control variates are a very well known way to reduce the variance of Monte Carlo estimates;
see e.g. the books of \cite{Robert2004}, \cite{GlassermanBook} and  \cite{Owen2013}.

Suppose we want to estimate the quantity $p(f) = \E_{p}[f(X)]$ for a suitable $f: \R^{d} \rightarrow \R$, based on an IID (independent and identically distributed) sample $\{X_{1}, \dots, X_{N}\}$ from distribution $p$. (The generalisation of control variates to MCMC will be discussed in Section \ref{sec:cubeThin}.)

The usual Monte Carlo estimate of $p(f)$ is 
\begin{equation}\label{eq:usualEstim}
\hat{p}(f) = \dfrac{1}{N}\sum_{n=1}^{N}f(X_{n}).
\end{equation}

Assume we know $J\in\mathbb{N}^\star$ functions $h_{j}: \R^{d} \rightarrow \R$ for $j\in\{1, \dots, J\}$ such that $p(h_j)=0$. Functions with this property are called control variates.  We can use this property to build an estimate with a lower variance: let's denote $h(X) = (h_{1}(X), \dots, h_{J}(X))^{t}$ and write our new estimate:
\begin{equation}\label{eq:controlVariatesEstimator}
    \hat{p}_{\beta}(f) = \dfrac{1}{N} \sum_{n=1}^{N}f(X_{n}) + \beta^{t}h(X_{n})
\end{equation}
with $\beta\in \R^{J}$. Then it is straightforward to show that $\E[\hat{p}_{\beta}(f)] = \E[\hat{p}(f)] = p(f)$. Depending on the choice of $\beta$ we may have $\Var[\hat{p}_{\beta}(f)] \leq \Var[\hat{p}(f)]$. The next section discusses how to choose such a $\beta$. 

%and the variance reduction we get using $\hat{p}_{\beta}(f)$ instead of $\hat{p}(f)$ depends on the choice of $h$. In general the more $h(X)$ is correlated with $f(X)$, the greater the variance reduction, see e.g \cite{Robert2004} \cite{Owen2013} and \cite{Dellaportas2011}. 

\subsection{Control variates as a weighting scheme}
\label{sub:cv_weights}

The standard approach to choose $\beta$ consists of two steps. 
First, one shows easily that  the value the minimises the variance of estimator \eqref{eq:controlVariatesEstimator} is: 
\begin{equation}\label{eq:betaOpt}
\beta^\star(f) = \Var(h(X))^{-1}\Cov(h(X), f(X))
\end{equation}
where $\Var(h(X))$ is the $J\times J$ variance matrix of the vector $h(X)$ and $\Cov(h(X), f(X))$ is the $J\times 1$ vector such that $\Cov(h(X), f(X))_{i, 1} = \Cov(f(X), h_{i}(X))$.

Second, one realises that this quantity may be estimated from the sample $X_1,\ldots, X_N$ through a simple linear regression model, where the $f(X_n)$'s are the outcome, and the $h_j(X_n)$'s are the predictors: 
\begin{equation}\label{eq:linRegression}
    f(X_{n}) \approx \mu + \beta^t h(X_{n}) + \epsilon_{n}, \quad \E[\epsilon_n] = 0.
\end{equation}

More precisely, let $\gamma\in R^{J+1}$ be the vector such that  $\gamma^t=(\mu, \beta^t)$, 
$H=(H_{ij})$ the design matrix such that $H_{i1}=1$, $H_{i(j+1)}=h_j(X_i)$,
and $F=(f(X_1),\ldots, f(X_N))$. Then the OLS (ordinary least squares) estimate of $\gamma$ is 
\begin{equation}\label{eq:ols}
\widehat{\gamma}_\mathrm{OLS} = (H^t H)^{-1} H^t F. 
\end{equation}

Since  $\E[f(X_n)]=\mu$ in this artificial regression model, 
the first component of $\widehat{\gamma}_{\mathrm{OLS}}$:
\begin{equation}\label{eq:optcvest}
\hat{p}_\star(f) \eqdef \widehat{\gamma}_\mathrm{OLS} \times e_1,
\end{equation}
actually corresponds to estimate \eqref{eq:controlVariatesEstimator} when  $\beta=\widehat{\beta}_\mathrm{OLS}$. 

At first glance, the approach described above seems to require implementing a different linear regression for each function $f$ of interest. 
\cite{Owen2013} noted however that one may re-express \eqref{eq:optcvest} as 
a weighted average: 
\begin{equation}\label{eq:cvest_as_wavg}
\hat{p}_\star(f) = \sum_{n=1}^N w_n f(X_n) 
\end{equation}
where the weights $w_n$ sum to one, and do not depend on $f$. It is thus possible to 
compute these weights once from a given sample (given a certain choice of control
variates), and then quickly compute $\hat{p}_\star(f)$ for any function $f$ of interest.

The exact expression of the weights are easily deduced from \eqref{eq:optcvest} and 
\eqref{eq:ols}: $w=(w_n)$ with 
\[
w = H(H^t H)^{-1} e_1.
\]

\subsection{Gradient-based control variates}\label{sec:steinCV}

In this section and the next, we recall generic methods to construct control variates. 
This section considers specifically control variates that derives from the score function, 
$s_p(x)=\nabla \log p(x)$. (We therefore assume that this quantity is tractable.) 

Under the following two conditions:
\begin{enumerate}
    \item the probability density $p \in C^{1}(\Omega, \
    R)$ where $\Omega\subseteq \R^{d}$ is an open set;
    
    \item Function $\phi\in C^{1}(\Omega, \R^{d})$ is such that  $\oint_{\partial\Omega} p(x)\phi(x)\cdot n(x) S(dx) = 0$ where $\oint_{\partial\Omega}$ denotes the integral over the boundary of $\Omega$, and $S(dx)$ is the surface element at $x\in\partial\Omega$; 
\end{enumerate}
the following function: 
\begin{equation}\label{eq:controlVariate}
    h(x) = \nabla_{x}\cdot \phi(x) + \phi(x)\cdot s_{p}(x)
\end{equation}
is a control variate:  $p(h)=  0$, see e.g. \cite{mira2013zero} or 
\cite{Oates2016} for further details. To get some intuition, note that in dimension 1 and assuming the domain of integration is an interval $]a,b[ \subset\R$, this amounts to an integration by part with the condition that $h(b)p(b) - h(a)p(a) = 0$.

Thus, whenever the score function is available (and the conditions above hold), we are able to construct an infinite number of control variates (one for each function $\phi$). 
For simplicity, we shall focus on the following  standard classes of such functions. 
First, for $i=1,\ldots, d$,
\begin{align*}
    \phi_{i} \colon \mathbb{R}^{d} &\rightarrow \mathbb{R}^{d}\\
    x  &\mapsto e_{i}
\end{align*}
which leads to the following  $d$ control variates: 
\begin{equation}\label{eq:scoreControlVariates}
    h_{i}(x) = s_{p}(x)[i].
\end{equation}
%We call these control variates the "score control variates".

For a Gaussian target, $N(\mu, \Sigma)$, the score is $s_p(x)=-\Sigma^{-1}(x-\mu)$, 
and the control variates above make it possible to reweigh the Monte Carlo sample
to make it have the same expectation as the target distribution. 

Second, we  consider, for $i,j=1,\ldots, d$:
\begin{align*}
    \phi_{ij} \colon \mathbb{R}^{d} &\rightarrow \mathbb{R}^{d}\\
    x  &\mapsto x[i]e_{j}
\end{align*}
which leads to the following $d^2$ control variates: %"mean control variates": 
\begin{equation}\label{eq:meanControlVariates}
    h_{ij}(x) = \ind\{i=j\} + x[i] s_{p}(x)[j].
\end{equation}
Again, for a Gaussian target $N(\mu, \Sigma)$, this makes it possible to fix
the empirical covariance matrix to true covariance $\Sigma$. 

In our simulations, we consider two sets of control variates: 
the `full' set, consisting of the $d$ control variates defined by 
\eqref{eq:scoreControlVariates}, and the $d^2$ control variates defined 
by \eqref{eq:meanControlVariates}. And a `diagonal' set of $2d$ control variates, 
where for \eqref{eq:meanControlVariates}, we only consider the cases where $i=j$.  
Of course, the former set should lead to better performance (lower variance), but 
since the complexity of our approach will be $\bigO(J^3)$, where
$J$ is the number of control variates, taking $J=\bigO(d^2)$ may be too expensive
whenever the dimension $d$ is large.

% Third,
% \begin{align*}
%     \phi_{ij} \colon \mathbb{R}^{d} &\rightarrow \mathbb{R}^{d}\\
%     x  &\mapsto x[i]x[j]e_{i}
% \end{align*}
% which leads to  $d^{2}$ "covariance control variates": 
% \begin{equation}\label{eq:covarianceControlVariates}
%   h_{ij}(x) = x[j] + x[i]x[j] s_{p}[i](x).  
% \end{equation}

\subsection{MCMC-based control variates}\label{sec:dellaportas}

We mention in passing other ways to construct control variates, in particular
in the context of MCMC. 

For instance, \cite{Dellaportas2011} noted that, for a  Markov chain $\{X_n\}$, the quantity 
\[
\phi(X_n) - \E\left[ \phi(X_n) | X_{n=1}\right]
\]
has expectation zero. In particular, if the MCMC kernel is a Gibbs sampler,
it is likely that one is able to compute the conditional expectation of 
each component; i.e. $\phi(x)=x[i]$ for $i=1,\ldots, d$. 

See also \cite{hammer2008control} for another way to construct control variates when the $X_n$'s are simulated from a Metropolis kernel. 

\section{The cube method}\label{sec:cube}

We review in this section the cube method of \cite{Deville2004}. This method originated from survey sampling, and is a way to sample from a finite population under constraints. The first subsection gives some definitions, the second one explains the flight phase of the cube method and the third subsection discusses the landing phase of the method.

\subsection{Definitions}

Suppose we have a finite population $\{1, \dots, N\}$ of $N$ individuals and that to 
each individual $n=1, \ldots, N$ is associated a variable of interest $y_{n}$ and $J$ 
auxiliary variables, $v_{n} = (v_{n1},  \dots, v_{nJ})$. 
Without loss of generality, suppose also that the $J$ vectors $(v_{1j}, \dots, v_{Nj})$ 
are linearly independent. We are interested in estimating the quantity 
$Y = \sum_{n=1}^N y_{n}$ using a  subsample of $\{1, \ldots, N\}$.  
Furthermore, we know the exact value of each sum $V_j=\sum_{n=1}^N v_{nj}$, 
and we wish to use this auxiliary information to better estimate $Y$. 

We assign, to each individual $n$,  a sampling probability $\pi_{n}\in[0, 1]$.
We consider random variables $S_n$ such that, marginally, $\Proba(S_{n} = 1) = \pi_{n}$. 
We may then define the Horvitz-Thompson estimator of $Y$:
\begin{equation}\label{eq:HorvitzThompsonEstim}
\hat{Y} = \sum_{n=1}^N \frac{S_n y_n}{\pi_{n}}
\end{equation}
which is unbiased, and which depends only on selected individuals (i.e $S_n=1$). 

We define similarly the Horvitz-Thompson estimator of $V_j$: 
\begin{equation}
    \hat{V}_{j} = \sum_{n=1}^N \frac{S_n v_{nj}}{\pi_n}.
\end{equation}

Our objective is to construct a joint distribution $\xi$ for the inclusion variables $S_n$ such that
$\Proba_\xi(S_n = 1)=\pi_n$ for all $n=1,\ldots, N$, and 
\begin{equation}\label{eq:balancing}
\hat{V} = V\quad\mbox{$\xi$-almost surely.}
\end{equation}
where $V=(V_1,\ldots, V_J)$, $\hat{V}=(\hat{V}_1,\ldots,\hat{V}_J)$. 
Such a probability distribution is called a balanced sampling design.

% Intuitively, we are trying to subsample $U$ while forcing some quantities to be equal for the subsampled population and the full population. Doing so, we can expect the subsampled population to be "representative" of the full population and thus the Horvitz-Thompson estimator (\ref{eq:HorvitzThompsonEstim}) to "approximate" $Y$ accurately.

\subsection{Subsamples as vertices}

We can view all the possible samples from $\{1, \ldots, N\}$ as the vertices of the hypercube $\mathcal{C}=[0, 1]^{N}$ in $\R^{N}$. A sampling design with inclusion probabilities $\pi_{n} = \Proba_\xi(S_n = 1)$ is then a distribution over the set of these vertices such that $\E[S] = \pi$, where $S=(S_1, \ldots, S_N)^t$, 
and $\pi=(\pi_1,\ldots, \pi_N)^t$ is the vector of inclusion probabilities. 
Hence, $\pi$ is expressed as a convex combination of the vertices of the hypercube.

We can think of a sampling algorithm as finding a way to reach any vertex of the cube, starting at $\pi$, while satisfying the balancing equation \eqref{eq:balancing}.
But before we describe such a sampling algorithm, we may wonder if it is possible to find a vertex such that \eqref{eq:balancing} is satisfied.

\subsection{Existence of a solution}

The balancing equation \eqref{eq:balancing} defines a linear system. Indeed, we can re-express \eqref{eq:balancing} as $S$ being a solution to $As=V$, where $A=(A_{jn})$ is of dimension $J\times N$, $A_{jn} = v_{kn} / \pi_n$.
This system defines a hyperplane $Q$ of dimension $N-J$ in $\R^{N}$.

What we want is to find vertices of the hypercube $\mathcal{C}$ that also belong to the hyperplane $Q$. Unfortunately, it is not necessarily possible, as it depends on how the hyperplane $Q$ intersects the cube $\mathcal{C}$. In addition, there is no way to know beforehand if such a vertex exists. Since $\pi \in Q$, we know that $\mathcal{K} \eqdef \mathcal{C}\cap Q \ne \emptyset$ and is of dimension $N-J$. The only thing we can say is stated Proposition 1 in \cite{Deville2004}: if $r$ is a vertex of $\mathcal{K}$, then in general $q=\mathrm{card}(\{n:\, 0 <r[n] < 1\}) \leq J$.
 
The next section describes the flight phase of the cube algorithm, which
generates a vertex in $\mathcal{K}$ when such vertices exist, or which, alternatively, 
returns a point in $\mathcal{K}$ with most (but not all) components set to zero or 
one. In the latter case, one needs to implement a landing phase, which is discussed
in Section \ref{sub:landing}.

\subsection{Flight phase}\label{sub:flight}

The flight phases simulates a process $\pi(t)$ which takes values in $\mathcal{K}=\mathcal{C}\cap Q$, and starts at $\pi(0)=\pi$. At every time $t$, one selects a unit vector $u(t)$, then one chooses randomly between one of the two points that are in the intersection of the hyper-cube $\mathcal{C}$ and 
the line parallel to $u(t)$ that passes through $\pi(t-1)$. The probability of selecting 
these two points are set to ensure that $\pi(t)$ is a martingale; in that way, we have $\E[\pi_t] = \pi$ at every time step. The random direction $u(t)$ must be generated to fulfil the following two requirements: 
(a) that the two points are in $Q$; i.e.  $u(t)\in \mathrm{ker} A$; 
and (b) whenever $\pi(t)$ has reached one of the faces of the hyper-cube, it must stay within that face; thus, $u(t)[k]=0$ if $\pi(t-1)[k]=0$ or $1$. 

Algorithm \ref{algo:flight} describes one step of the flight phase. 

\begin{algorithm}\label{algo:flight}
\caption{Flight phase iteration}
    \KwInput{$\pi(t-1)$}
    \KwOutput{$\pi(t)$}
    Sample $u(t)$ in $\ker A$ with $u_{k}(t) = 0$ if the $k$-th component of $\pi(t-1)$ is an integer.
    
    Compute $\lambda^{\star}_{1}$ and $\lambda^{\star}_{2}$, the largest values of $\lambda_{1} > 0$ and $\lambda_{2} > 0$ such that: $0\leq \pi(t-1) + \lambda_{1}u(t) \leq 1$ and $0\leq \pi(t-1) - \lambda_{2} u(t) \leq 1$.

    With probability $\lambda_2^\star / (\lambda_1^\star + \lambda_2^\star)$,
    set $\pi(t) \leftarrow \pi(t-1) + \lambda_1 u(t) $; 
    otherwise, set $\pi(t) \leftarrow \pi(t-1) - \lambda_2 u(t)$.

    % Select $\pi(t) = \pi(t-1) + \lambda_{1}(t)u(t)$ with probability $q(t) = \lambda^{*}_{2}(t)/(\lambda^{*}_{2}(t) + \lambda^{*}_{1}(t))$ otherwise select $\pi(t) = \pi(t-1) - \lambda_{2}(t)u(t)$
\end{algorithm}

The flight phase stops when Step 1 of Algorithm \ref{algo:flight} cannot be performed
(i.e. no vector $u(t)$ fulfils these conditions). Until this happens, each iteration 
increases by at least one the number of components in $\pi(t)$ that are either zero or one. Thus, the flight phases completes at most in $N$ steps. 

In practice, to generate $u(t)$, one may proceed as follows: first generate a random vector $v(t)\in \R^{N}$, then project it in the constraint hyperplane: $u(t) = I(t)v(t) - I(t)A^{t}(AI(t)A^{t})^{-}AI(t)v(t)$ where $I(t)$ is a diagonal matrix such that $I_{kk}(t)$ is 0 if $\pi_{k}(t)$ is an integer and 1 otherwise, and $M^{-}$ denotes the pseudo-inverse of the matrix $M$.

\cite{Chauvet2006} propose a particular method to generate vector $v(t)$ which ensures
that the complexity of a single iteration of the flight phase is $\bigO(J^3)$. This 
leads to an overall complexity of $\bigO(NJ^3)$ for the flight phase, since it terminates
in at most $N$ iterations. 

\subsection{Landing phase}\label{sub:landing}

Denote by $\pi^\star$ the value of process $\pi(t)$ when the flight phase terminates. 
If $\pi^\star$ is a vertex of $\mathcal{C}$ (i.e. all its components are either zero or one),
one may stop and return $\pi^\star$ as the output of the cube algorithm. 
If $\pi^\star$ is not a vertex, this informs us that no vertex
belongs to $\mathcal{K}$. One may implement a landing phase, which aims at choosing
randomly a vertex which is close to $\pi^\star$, and such that the variance of
the components of $\hat{V}$ is small. 

Appendix \ref{app:landing} gives more details on the landing phase. Note that its worst-case complexity is $\bigO(2^J)$. However, in practice, it is typically either much faster, or not required
(i.e. $\pi^\star$ is already a vertex) as soon as $J\ll N$.

\section{Cube thinning}
\label{sec:cubeThin}

%\subsection{General idea}\label{sub:cube_idea}

We now explain how the previous ingredients (control variates, and the cube method) may be
combined in order to  thin  a Markov chain, $X_1,\ldots, X_N$, into a sub-sample of size $M$. As before, the invariant distribution of the chain is denoted by $p$, and we assume we know of $J$ control variates $h_j$, i.e. $p(h_j)=0$ for $j=1,\ldots,J$. 

\subsection{First step: computing the weights}

The first step of our method is to use the $J$ control variates to compute the $N$ weights
$w_n$, as defined at the end of Section \ref{sub:cv_weights}. Recall that these weights sum to one, that they  automatically fulfil the constraints:
\begin{equation}\label{eq:constrainthj}
\sum_{n=1}^N w_n h_j(X_n) = 0
\end{equation}
for $j=1, \ldots, J$, and that we use them to compute
\begin{equation}\label{eq:ourest}
\hat{p}_\star(f) = \sum_{n=1}^N w_n f(X_n)
\end{equation}
as a low-variance estimate for $p(f)$ for any $f$. 

Recall that the control variates procedure we described in Section 
\ref{sec:controlVariates} 
assume that the input variables, $X_1,\ldots, X_N$, are  IID. 
This is obviously not the case in a MCMC context; however, we follow the common practice 
\citep{mira2013zero, Oates2016} of applying the procedure to MCMC points as if they were IID points. This implies that the weighted estimate above corresponds to a value of $\beta$
in \eqref{eq:controlVariatesEstimator} that does not minimise the (asymptotic) variance of estimator \eqref{eq:controlVariatesEstimator}. 
It is actually possible to estimate the value of
$\beta$ that minimises the asymptotic variance of a MCMC estimate 
\citep{Dellaportas2011, brosse2019diffusion}. 
However, this type of approach is specific to certain MCMC samplers, and, critically for us, it cannot be cast as a weighting scheme. Thus we stick
to this standard approach. 

We note in passing that, in our experiments (see Figure \ref{fig:lotka:weights} and the 
surrounding discussion) the weights $w_n$ makes it easy to assess visually 
the convergence (and thus the burn-in) of the Markov chain. In fact, since the MCMC 
points of the burn-in phase are far from the mass of the target distribution, the procedure must assign  a small or negative weight to these points in order to respect the constraints
based on the control variates. Again, see Section \ref{sub:lotka} for more discussion 
on this issue. 
The fact that control variates may be used to assess MCMC convergence has been known
for a long time \citep[e.g.][]{brooks1998some}, but the visualisation of weights 
makes this idea more expedient.

\subsection{Second step: cube resampling} 

The second step consists in resampling the weighted sample $(w_n, X_n)_{n=1,\ldots,N}$, 
to obtain a sub-sample $\mathcal{S}=\{X_n:\, S_n=1\}$
where $S_n$ are random variables such that 
(a) $\E[S_n]=w_n$; 
(b) $\sum_{n=1}^N S_n=M$, and (c) for $j=1,\ldots, J$:  
\[ 
\sum_{S_n=1} h_j(X_n) = 0. 
\]

Condition (a) ensures that the procedure does not introduce 
any bias:
\[
\E\left[ \frac 1 M \sum_{S_n=1} f(X_n) \bigg\rvert X_{1:N} \right] 
= \sum_{n=1}^N w_n f(X_n). 
\]
Condition (b) ensures that the sub-sample is exactly of size $M$. 

We would like to use the cube method in order to generate the $S_n$'s. Specifically, we would like to assign the inclusion probabilities $\pi_n$ 
to $w_n$, and impose the $(J+1)$ constraints defined above by Conditions (b) and (c). 
There is one caveat, however: the weights $w_n$ do not necessarily lie in $[0,1]$. 

\subsection{Dealing with weights outside of $[0, 1]$}
\label{sub:negative_weights}

We  rewrite \eqref{eq:ourest} as:
\begin{equation}
    \hat{p}_\star(f) 
    %= \sum_{n=1}^{N} \sgn(w_{n})|w_{n}|f(X_{n}) 
    =\dfrac{\Omega}{M} \times \sum_{n=1}^{N} W_n \times \sgn(w_{n}) f(X_{n})
\end{equation}
where $\Omega = M^{-1} \sum_{n=1}^{N}|w_{n}|$ and $W_{n} = M  |w_{n}| / \Omega$. 
We now have $W_n\geq 0$, and $\sum_{n=1}^N W_n = M$, which is required for condition (b)
in the previous section. We might have a few points such that $W_n>1$.  In that case, we  replace them by $\lfloor W_n \rfloor$ copies, with adjusted weights $W_n / \lfloor W_n \rfloor$. 

It then becomes possible to implement the cube method, using as inclusion probabilities
the $W_n$'s, and as the matrix $A$ that defines the $J+1$ constraints, the matrix 
$A=(A_{jn})$ such that $A_{1n}=1$, $A_{(j+1)n} = \sgn(w_n) h_j(X_n) $. The cube method 
samples variables $S_n$, which may be used to compute the sub-sampled estimate
\begin{equation}\label{eq:final_estimate}
\hat{\nu}(f) = \frac{\Omega}{M} \sum_{S_n=1}\sgn(w_{n})f(X_{n}). 
\end{equation}

More generally, in our numerical experiments, we shall evaluate to which extent
the random signed measure: 
\begin{equation}\label{eq:measure}
\hat{\nu} = \frac{\Omega}{M} \sum_{S_n=1}\sgn(w_{n})\delta_{X_n}(\mathrm{d} x). 
\end{equation}
is a good approximation of the target distribution $p$.

\section{Experiments}\label{sec:experiments}

We consider two examples. The first example is taken from \cite{Riabiz2020}, and is 
used to compare cube thinning with KSD thinning. The second example illustrates cube
thinning when used in conjunction with control variates that are not gradient-based. 
We also include standard thinning in our comparisons. 

Note that there is little point in comparing these methods in terms of CPU cost, 
as KSD thinning is considerably slower than cube thinning and standard thinning whenever $M\gg 100$. (In one of our experiment, for $M=1000$, KSD took close to 7 hours to run, while cube thinning with all the covariates took about 30 seconds.)  
Thus, our comparison will be in terms of statistical error, or, more precisely, 
in terms of how representative of $p$ is the selected sub-sample. 

% We discussed three different ways of thinning: our cube method, the KSD thinning method of  and the regular thinning method. Among these methods, the KSD method is much slower as soon as we take $M\gg 100$. Thus, it does not make sense to compare these methods based on their CPU time. Instead, we only seek to compare their statistical properties. More precisely, we compare these three algorithms in terms of how well their output "represent" the target distribution in Section~\ref{sec:lotka}. We will elaborate on the meaning of "representative" in the next section. 

% We also want to assess the performances of the cube method with control variates when the gradient of the log-density is not available. That's what we do in Section~\ref{sec:truncNorm} where we compare the variances of the cube estimator (\ref{eq:resampEstim}), the usual thinning estimator, the regression estimator (\ref{eq:newEstimator}) and the usual MCMC estimator.

In the following (in particular in the plots), "cubeFull" (resp. "cubeDiagonal") will refer to our approach based on the full (resp. diagonal) set of control variates, as discussed in 
Section \ref{sec:steinCV}. The mention "NoBurnin" means that burn-in has been 
discarded manually (hence no burn-in in the inputs).  Finally, "thinning" denotes the usual thinning approach, "SMPCOV", "MED" and "SCLMED" are the same names used in \cite{Riabiz2020}
for KSD thinning, based on three different kernels.

To implement the cube method, we used R package \texttt{BalancedSampling}.

\subsection{Evaluation criteria}

We could compare the three different methods in terms of variance of the estimates of 
$p(f)$ for certain functions $f$. However, it is easy to pick functions $f$ that
are strongly correlated with the chosen control variates; that would bias the comparison
in favour of our approach.  In fact, as soon as the target is Gaussian-like, the control variates we chose in Section \ref{sec:steinCV} should be strongly correlated with the
expectation of any polynomial function of order two, as we discussed in that section.
 
Rather, we consider criteria that are indicative of the performance of the methods for a general class of function. Specifically, we consider three such criteria. 
The first one is the kernel Stein discrepency (KSD) as defined in \cite{Riabiz2020} and recalled in the introduction, see \eqref{eq:ksd}. Note that this criterion is particularly favourable to KSD thinning, since this approach specifically minimises this quantity. 
(We use the particular version based on the median kernel in \citet{Riabiz2020}.)

The second criterion is the energy distance (ED) between $p$ and the empirical distribution defined by the thinning method; e.g. \eqref{eq:measure} for cube thinning.
Recall that the ED between two distributions $F$ and $G$ is:  
\begin{equation}\label{eq:basicED}
    ED(F, G) = 2\E||Z - X||_{2} - \E||Z - Z'||_{2} - \E||X - X'||_{2}
\end{equation}
where $Z', Z\overset{iid}{\sim} F$ and $X', X\overset{iid}{\sim} G$,
and that this quantity is actually a pseudo-distance: $ED(F, G)\geq 0$, $ED(F, G)=0 \Rightarrow F=G$, $ED(F, G) = ED(G, F)$, but ED does not fulfil the triangle inequality
 \citep{Szekely2005, Klebanov2006}.

One technical difficulty is that \eqref{eq:measure} is a signed measure, not a probability measure; see Appendix \ref{app:ed} on how we dealt with this issue. 

Our third criteria is inspired by the star discrepancy, a well-known measure of 
the uniformity of $N$ points $u_n\in[0,1]^d$ in the context of quasi-Monte Carlo 
sampling  \citep[][Chap. 15]{Owen2013}. 
Specifically, we consider the quantity 
\[
d^\star(\hat{P}, \hat{\nu}) = \sup_{B\in\mathcal{B}} 
\left| \hat{P}_\psi(B) - \hat{\nu}_\psi(B) \right|
\]
where $\psi:\R^d \rightarrow [0, 1]^d$, $\hat{P}_\psi$ and $\hat{\nu}_\psi$ 
are the push-forward measures associated to empirical distributions 
$\hat{P}=(N-b)^{-1}\sum_{n=b+1}^N \delta_{X_n}(dx)$, and $\hat{\nu}$ as defined in 
\eqref{eq:measure}, and 
$\mathcal{B}$ is the set of hyper-rectangles $B=\prod_{i=1}^d [0, b_i]$.
In practice, we defined function $\psi$ as follows: we apply the linear transform
that makes the considered sample to have zero mean and unit variance, and then we 
applied the inverse  CDF (cumulative distribution function) of a unit Gaussian
to each component. 

Also, since the sup above is not tractable,
we replace it  by a maximum over a finite number of $b_i$ (simulated uniformly).

\subsection{Lotka-Volterra model}\label{sub:lotka}

This example is taken from \cite{Riabiz2020}.
The Lotka-Volterra model describes the evolution of a prey-predator system in a closed environment. We denote the number of prey by $u_{1}$ and the number of predator by $u_{2}$. The growth rate of the prey is controlled by a parameter $\theta_{1} > 0$ and its death rate - due to the interactions with the predators - is controlled by a parameter $\theta_{2} > 0$. In the same way, the predator population has a death rate of $\theta_{3} > 0$ and a growth rate of $\theta_{4} > 0$. Given these parameters, the evolution of the system is described by a system of ODEs:
\begin{eqnarray*}
\dfrac{du_{1}}{dt} = & \theta_{1}u_{1} - \theta_{2}u_{1}u_{2}\\
\dfrac{du_{2}}{dt} = & \theta_{4}u_{1}u_{2} - \theta_{3}u_{2}
\end{eqnarray*}

\cite{Riabiz2020} set $\theta = (\theta_{1}, \theta_{2}, \theta_{3}, \theta_{4})= (0.67, 1.33, 1, 1)$, the initial condition $u_{0} = (1, 1)$, and simulate synthetic data.
They assume they  observe the populations of prey and predator at  times $t_{i}, i = 1, \dots, 2400$ where the $t_{i}$ are taken uniformly on $[0, 25]$ and that these observations are corrupted with a centered Gaussian noise with a covariance matrix $C = \mbox{diag}(0.2^{2}, 0.2^{2})$.
Finally, the model is parametrized in terms of $x =(\log\theta_{1}, \log\theta_{2}, \log\theta_{3}, \log\theta_{4}) \in \R^{4} $ 
and a standard normal distribution as a prior on $x$ is used.

The authors have provided their code as well as the sampled values they got by running different MCMC chains for a long time. We use the exact same experimental set-up, and we do not run any MCMC chain on our own, but use the ones they provide instead; specifically 
the simulated chain, of length $2\times 10^6$, from  preconditionned-MALA. 

% Since our control variates require the availability of the gradient of the target, we assume we are able to run a MCMC algorithm taking the gradient information into account and we decide to run our experiments on the P-MALA chain produced using a preconditionned-MALA algorithm provided by \cite{Riabiz2020}. This chain is of length $2 \times 10^{6}$ and 

% We take a burnin period of 2000 iterations for the cube method based on the visual inspection of the weights - see below - and a burnin period of 90000 iterations for the thinning method based on the value given in \cite{Riabiz2020}.

\begin{figure}
    \centering
    \includegraphics[width=\linewidth]{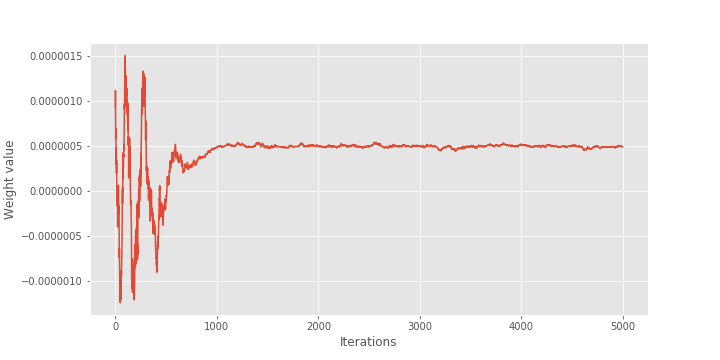}
    \includegraphics[width=\linewidth]{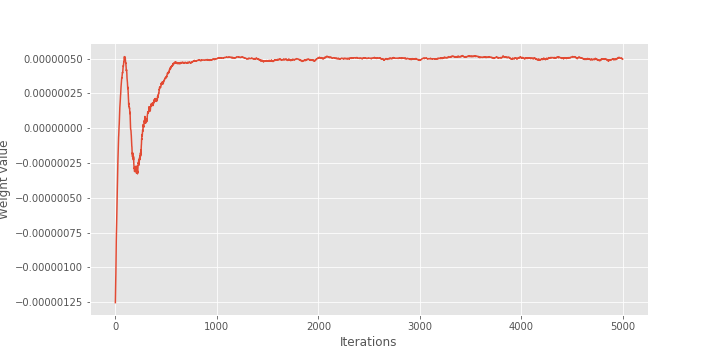}
    \caption{Lotka-Volterra example: first 5000 weights of the cube methods, 
    based on full (top) or diagonal (bottom) set of covariates.}%
    \label{fig:lotka:weights}
\end{figure}

We  compress this chain into a subsample of size either $M=100$ or  $M=1000$. For each value of $M$, we run different variations of our cube method   50 times and make a comparison with the usual thinning method and with the KSD thinning method with different kernels, see \cite{Riabiz2020}. 
In Figure \ref{fig:lotka:weights} we show the first 5000 weights of the cube method. We can see that after 1000 iterations, the weights seem to stabilize. Based on visual examination of these weights, we choose a conservative burnin period of 2000 iterations for the variants 
where burn-in is removed manually.

\begin{figure}
    \centering
    \includegraphics[width=\linewidth]{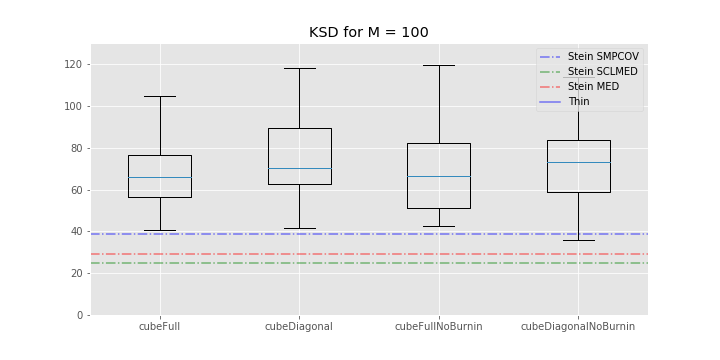}
    \includegraphics[width=\linewidth]{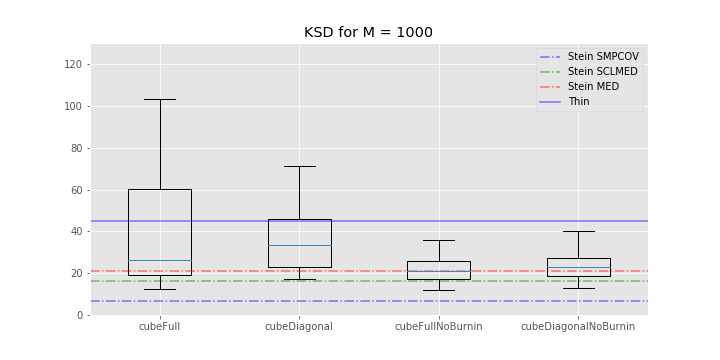}
    \caption{Lotka-Volterra example: box-plots of the kernel Stein discrepency for all the cube method variations, the KSD method for three kernels and the usual thinning method. Top: $M=100$. Bottom: $M=1000$. (In the top plot, standard thinning is omitted to improve
    clarity, as corresponding value is too high.)}%
    \label{fig:lotka:KSD}
\end{figure}

\begin{figure}
    \centering
    \includegraphics[width=\linewidth]{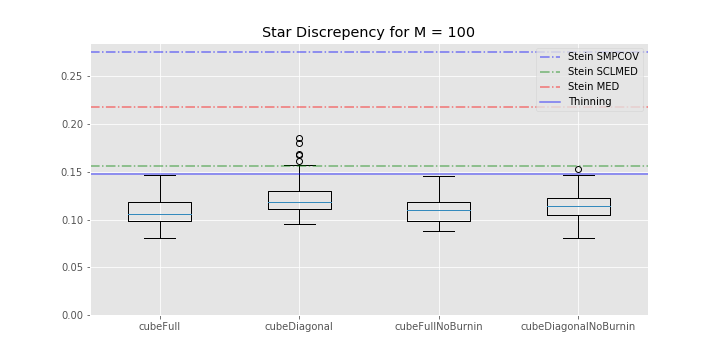}
    \includegraphics[width=\linewidth]{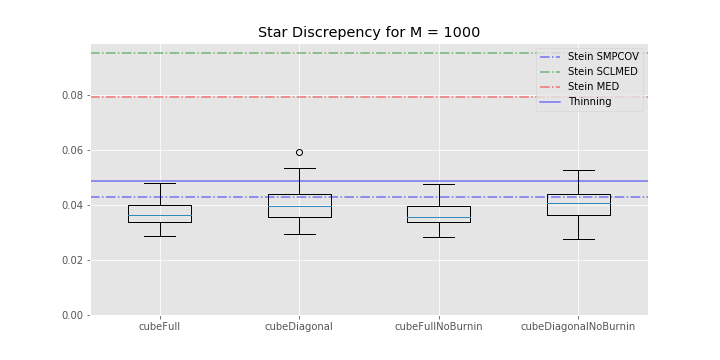}
    \caption{Lotka-Volterra example: box-plots of the star discrepency for all the cube method variations, the KSD method for three kernels and the usual thinning method. Top: $M=100$. Bottom: $M=1000$.}%
    \label{fig:lotka:starDiscrepency}
\end{figure}

\begin{figure}
    \centering
    \includegraphics[width=\linewidth]{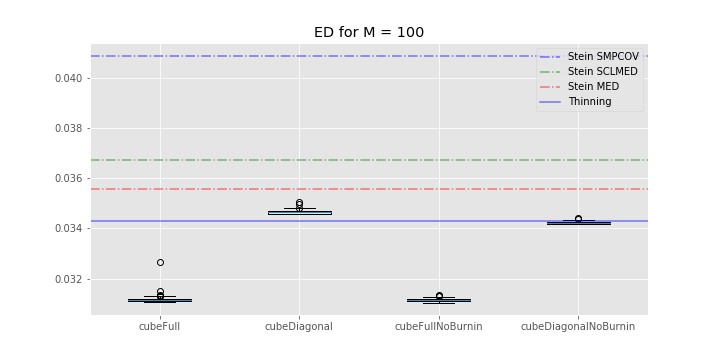}
    \includegraphics[width=\linewidth]{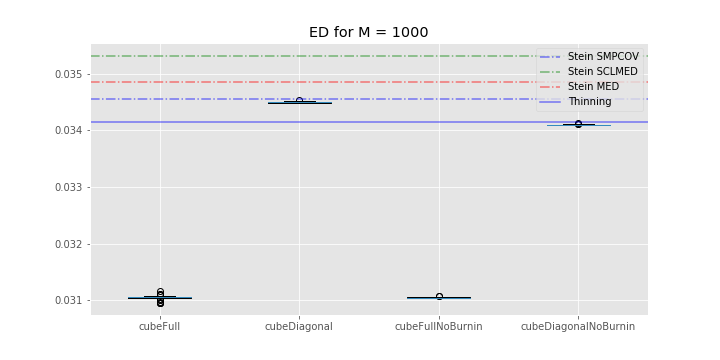}
    \caption{Lotka-Volterra example: boxplots of the energy distance for all the cube method variations, the KSD method for three kernels and the usual thinning method. Top: $M=100$. Bottom: $M=1000$.}%
    \label{fig:lotka:EnergyDistance}
\end{figure}

We plot the results of the experiment on Figures
\ref{fig:lotka:starDiscrepency}, \ref{fig:lotka:KSD} and \ref{fig:lotka:EnergyDistance}.

First, we see that regarding the kernel Stein discrepancy metric, Figure \ref{fig:lotka:KSD}, the KSD method performs better than the standard thinning procedure and the cube method. This is not surprising since even if this method does not properly minimizes the Kernel-Stein Discrepency, this is still its target. We also see that for $M=1000$, the KSD method performs a bit better  than our cube method which in turn performs better than the standard thinning procedure. Note that the relative performance of the KSD method to our cube methods depends on the kernel that is being used and that there is no way to determine which kernel will perform best before running any experiment. 

The picture is different for $M=100$:  KSD thinning outperforms standard thinning, which in turn outperforms all of our cube thinning variations.  Once again, the fact that the KSD method performs better than any other method seems reasonable: since it is about minimizing the Kernel-Stein Discrepancy, the KSD method is "playing at home" on this metric. 

If we look at Figure \ref{fig:lotka:EnergyDistance}, we see that all of our cube methods outperform the KSD method with any kernel. Interestingly, the standard thinning methods has a similar Energy Distance as the cube methods with "diagonal" control variates. These observations are true for both $M=100$ and $M=1000$. We can also note that the cube method with the full set of control variates tends to perform much better than its "diagonal" counterpart, whatever the value of $M$.

Finally, looking at Figure \ref{fig:lotka:starDiscrepency}, it is clear that the KSD method - with any kernel - performs worse than any cube method in terms of star discrepancy.

Overall, the relative performance of the cube methods and KSD methods can change a lot depending on the metric being used and the number of points we keep. In addition, while all the cube methods tend to perform roughly the same, this is not the case of the KSD method, whose performances depend on the kernel we use. Unfortunately, we have no way to determine beforehand which kernel will perform best. This is a problem since the KSD method is computationally expensive for subsamples of cardinal $M\gg 100$. 

Thus, by and large, cube thinning seems much more convenient to use (both in terms
of CPU time and sensitivity to tuning parameters) while offering, roughly, the same level of 
statistical performance.

%On the contrary, the cube method scales very well with the number of subsampled points.

\subsection{Truncated Normal}\label{sec:truncNorm}

In this example, we use the (random-scan version of) the Gibbs sampler of \cite{Robert2004} to sample from 10-dimensional multivariate normal truncated to $[0, \infty)^{10}$.
We generated the parameters of this truncated normal as follows: the mean was set as the 
realization of a 10-dimensional standard normal distribution, while for the covariance 
matrix $\Sigma$ we first generated a matrix $M\in \mathcal{M}_{10,10}(\R)$ for which each 
entry was the realization of a standard normal distribution. Then we set $\Sigma = M^{T}M$.

Since we are using a Gibbs sampler, we have access to the Gibbs control variates of 
 \cite{Dellaportas2011}, based on the expectation of each update (which amounts to
 simulating from a univariate Gaussian). Thus, we consider 10 control variates. 

% This time, we do not rely on the Stein control variates described in Section~\ref{sec:steinCV}. Instead, we follow and use the control variates defined in (\ref{eq:gibbsControlVariates}).

% In our case of a truncated normal distribution, we can easily compute every conditional mean and thus, in our example, we have 10 different control variates, one per conditional 
%mean.

The Gibbs sampler is run for $N=10^5$ iterations;  no burn-in is performed.  
We compare the following estimators of the expectation of the target distribution
the standard estimator, based on the whole chain (`usualEstim' in the plots), the  estimator based on standard thinning (`thinEstim' in the plots), the control variate estimator based on the whole chain, i.e. \eqref{eq:optcvest} ('regressionEstim' in the plots), 
and finally our cube estimator described in Section~\ref{sec:cubeThin} (`cubeEstim' in the plots). 
For standard thinning and cube thinning, the thinning sample size is set to $M=100$,
which corresponds to a compression factor of $10^3$.

% We make the following experiment: we run a random-scan Gibbs sampler, see \cite{Robert2004}, for $10^5$ steps, and we estimate the mean of the truncated normal distribution without removing any burnin. Suppose we get a sample $\{X_{1}, \dots, X_{N}\}$ produced by the random-scan Gibbs sampler. We compare several estimators:

\begin{figure}
    \centering
    \includegraphics[width=0.45\linewidth]{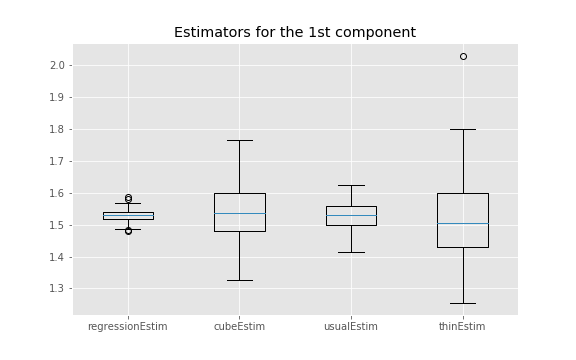}    \includegraphics[width=0.45\linewidth]{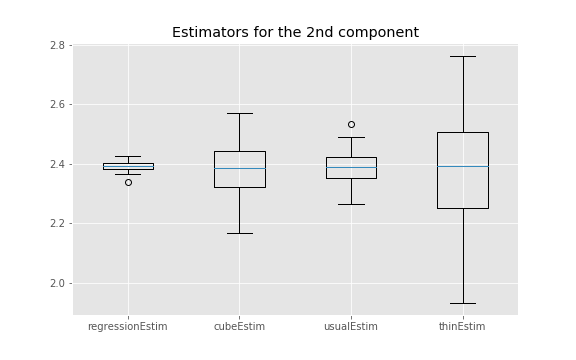}
    \includegraphics[width=0.45\linewidth]{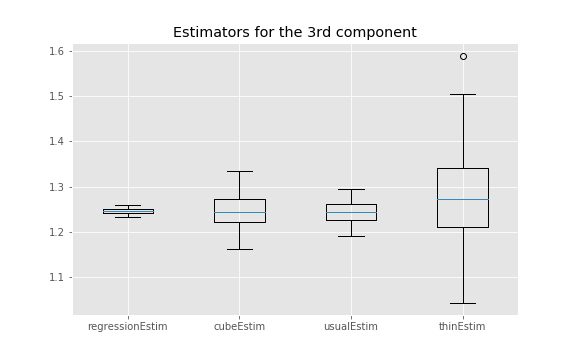}
    \includegraphics[width=0.45\linewidth]{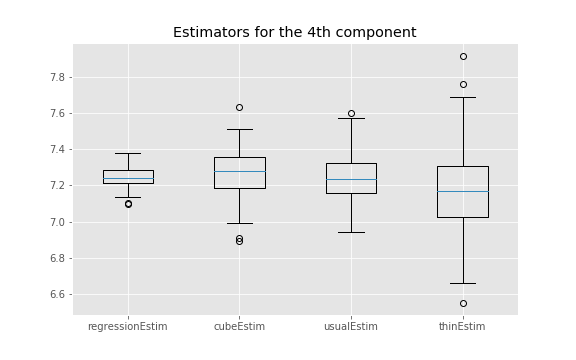}
    \caption{Truncated normal example:  box-plots over 100 independent replicates 
    of  each estimator; see text for more details.
\label{fig:truncNorm:boxplots}
}
\end{figure}

The results are shown in Figure \ref{fig:truncNorm:boxplots}. First, we can see that the control variates we chose lead to a substantial decrease in the variance of the estimates for regressionEstim compared to usualEstim. 
Second, the cube estimator performs worse than the regression estimator in terms of variance, but this was expected, as explained in Section~\ref{sec:cubeThin}. More interestingly, if we cannot say that the cube estimator performs better than the usual MCMC estimator in general, we can see that on some components it performs as good or even better, even though the cube estimator uses only $M=100$ points while the usual estimator uses $10^5$ points. This is largely due to the excellent choice of the control variates. Finally, the cube estimator outperforms the regular thinning estimator on every component, sometimes significantly.

\bibliography{cubeThinning} 
\bibliographystyle{apalike}

\appendix 

\section{Details on the landing phase}
\label{app:landing}

The landing phase seeks to generate a random vector $S$ in $\{0, 1\}^N$, with
expectation $\pi^\star$ (the output of the flight phase), which minimises the 
criterion $\mbox{tr}(M\Var(\hat{V}|\pi^\star))$ for a certain matrix $M$. 
(The notation $\cdot|\pi^\star$ refers to the distribution of $S$ conditional 
on $\pi(t)=\pi^\star$ at the end of the flight phase.) 

Since  $\Var(S) = \Var(\E[S|\pi^\star]) + \E[\Var(S|\pi^\star)]$ by the law of total variance,
and since the first term is zero (as $\E[S|\pi^\star] = \pi^\star$),
we have 
%and since  $A\Var(\pi^\star) A^{t} = 0$ as $\pi^\star\in Q$, we have:
\begin{equation}
    \Var(\hat{V}) = \E[\Var(\hat{V}|\pi^\star)] = \E[A\Var(S|\pi^\star)A^{t}] .
\end{equation}
and thus: 
\begin{equation}
    \mbox{tr}(M\Var(\hat{V}|\pi^\star)) = 
    \sum_{s\in \{0, 1\}^N} p(s|\pi^\star)(s-\pi^\star)^{t}A^{t}MA(s-\pi^\star). 
\end{equation}

Choosing $M = (AA^{t})^{-1}$, as recommended by \cite{Deville2004}, 
amounts to  minimising the distance to the hyperplane `on average'.
Let $C(s) =(s- \pi^\star)^{t} A^t(AA^t)^{-1}A^t (s- \pi^\star)$, then 
the minimisation program  is equivalent to the following linear programming 
problem over $q$ variables only:

\begin{equation}
    \min_{\xi^\star(.)} \sum_{s^\star\in\mathcal{S^\star}}C(s^\star)\xi^\star(s^\star)
\end{equation}
with constraints $\sum_{s^\star\in\mathcal{S^\star}} \xi^\star(s^\star) = 1$, $0\leq \xi^\star(s^\star) \leq 1$, $\sum_{s^\star\in\mathcal{S^\star}| s^\star_{k} = 1} \xi^\star(s^\star) = \pi^\star_{k}$ for every $k\in U^\star$ and $\mathcal{S^\star} = \{0, 1\}^{q}$ where $q = \mathrm{card}(U^\star)$ and $U^\star = \{k\in U:\, 0 < \pi^\star[k] < 1\}$. Here $\xi^\star$ denotes the marginal distribution of the components $U^\star$ of the sampling design $\xi$ and $C(s^\star)$ must be understood as $C(s)$ with the components of $s\notin U^\star$ being fixed by the result of flight phase, thus in this minimization problem $C$ is in fact depending on the components of $s$ that are in $U^\star$ only.

The constraints define a bounded polyhedron. By the fundamental theorem of linear programming, this optimization problem has at least one solution on a minimal support, see \cite{Deville2004}.

The flight phase ends on a vertex of $\mathcal{K}$ and, by Proposition 1 in \cite{Deville2004},  $q\leq J$; typically $J\ll N$. This means that we are solving a linear programming problem in a dimension $q$ potentially much lower than the population size $N$, and if we do not have too many auxiliary variables, this optimization problem will not be 
computationally too expensive. In practice, a simplex algorithm is used to find the solution.

% \subsection{Complexity of the cube method}

% TODO

% A quick calculation shows that the computational cost of one iteration of Algorithm \ref{algo:flight} is $\mathcal{O}(NJ^{2} + J^{3} + N^{2})$. Since we are doing at most $N$ iterations and at best $N-J$ iterations for the flight phase, the cost of this phase is $\mathcal{O}(N^{2}J^{2} + NJ^{3} + N^{3})$. In addition, the cost of the landing phase is at worst $\mathcal{O}(2^{J})$. Overall, the method's cost is $\mathcal{O}(N^{2}J^{2} + NJ^{3} + N^{3} + 2^{J})$ in the worst case.

% If we have only a handful of auxiliary variables, the landing phase will not be too much of a problem, in spite of the complexity being exponential in $J$. However, the polynomial dependence on the population size $N^{3}$ may prevent us from using this method in the context of thinning: it is only worth thinning a long chain - $N$ being large.

% Fortunately, the authors of \cite{Deville2004} devised an alternative and faster way to perform the flight phase in \cite{Chauvet2006}. In the case, we do not have to perform the classic flight phase described in \cite{Deville2004} at the end of the new and faster flight phase, the cost is $\mathcal{O}(NJ^{3})$. Which means that our cube algorithm depends only linearly on the population size.

\section{Estimation of the energy distance}
\label{app:ed}

There are two difficulties with computing  (\ref{eq:basicED}). First, it involves
intractable expectations. Second, as pointed out at the end of Section \ref{sub:negative_weights},  the empirical distribution
generated by cube thinning, \eqref{eq:measure}, is actually a signed measure. 

Regarding the first issue, we can  approximate (\ref{eq:basicED}) from our MCMC sample $X_{1}, \dots, X_{N}$. That is, if our subsampled empirical measure writes $\hat{\nu} = \sum_{m=1}^{M}w_{m}\delta_{Z_{m}}$ and that we approximate the distribution associated with $p$ by $\hat{P} = (N-b)^{-1}\sum_{n=b+1}^{N}\delta_{X_{n}}$ where $1\leq b\leq N$ is the burn-in of the chain, then, we can estimate $ED(\hat{\mu}, p)$ with $ED(\hat{\mu}, \hat{P})$.

Regarding the second issue, we can generalize the energy distance to finite measures: suppose we have two finite and potentially signed measures $\nu_{1}$ and $\nu_{2}$, both defined on the same measurable space $(\Omega, \mathcal{P}(\Omega)\}$ where $\Omega = \{X_{1}, \dots, X_{N}\}$ and $\mathcal{P}(\Omega)$ denotes the set of parts of $\Omega$. Suppose in addition that $\nu_{1}(\Omega) = \alpha_{1}$ and $\nu_{2}(\Omega) = \alpha_{2}$ with $\alpha_{1} \neq 0$ and $\alpha_{2} \neq 0$. We define the generalized energy distance as:
\begin{align*}
    ED^\star(\nu_{1}, \nu_{2}) =
    &
    \dfrac{2}{\alpha_{1}\alpha_{2}}\int_{\Omega} ||x - y||_{2}d\nu_{1}(x)d\nu_{2}(y) \\
    & - \dfrac{1}{\alpha_{1}^{2}}\int_{\Omega}||x - x'||_{2}d\nu_{1}(x)d\nu_{1}(x')  \\
    & - \dfrac{1}{\alpha_{2}^{2}}\int_{\Omega}||y - y'||_{2}d\nu_{2}(y)d\nu_{2}(y').
\end{align*}
Then, by negative definiteness of the application $\phi(x,y)=||x-y||_{2}$ on $\R^{N}\times\R^{N}$, we have that $ED^\star(\nu_{1}, \nu_{2}) \geq 0$ with equality if and only if $\dfrac{1}{\alpha_{1}}\nu_{1} = \dfrac{1}{\alpha_{2}}\nu_{2}$. Which means that the generalized energy distance is zero if and only if the two measures are equal up to a non-zero multiplicative constant, see \cite{Szekely2005} for a demonstration. This generalized energy distance is also symmetric, but the triangle inequality does not hold. It is a pseudo-distance.

Thus we will use the following criterion, which we will abusively call the energy distance in the rest of the paper:
\begin{align*}\label{eq:ed}
    {ED}^\star(\hat{\nu}, \hat{P}) =& \dfrac{2}{(N-b)\alpha_{1}}\sum_{k=1}^{N}\sum_{n=b+1}^{N}\dfrac{\Omega}{M}sgn(w_{k})||X_{k} - X_{n}||_{2} \mathbf{1}_{\{S_{k}=1\}}\\
    & -\dfrac{1}{\alpha_{1}^{2}}\sum_{n=1}^{N}\sum_{k=1}^{N}\left(\dfrac{\Omega}{M}\right)^{2} sgn(w_{n})sgn(w_{k}) ||Z_{k} - Z_{n}||_{2} \mathbf{1}_{\{S_{k}=1\}}\mathbf{1}_{\{S_{n}=1\}}
\end{align*} 
where $\hat{\nu}$ is defined in (\ref{eq:measure}) and we dropped the last term because it does not depend on $\hat{\nu}$ and it is a potentially expensive sum of $(N-b)^{2}$ terms.

Note that the probability of $\hat{\nu}(\Omega)$ being zero is non-null and then there is a non-negligible probability of ${ED}^\star(\hat{\nu}, \hat{P})$ being undefined. However, this event is unlikely to happen. %, and we stick with this criterion.

\end{document}